\documentclass[final,5p]{elsarticle}

\usepackage{graphicx}% Include figure files

\journal{Physics Letter B} % do not change

\begin{document}

\begin{frontmatter}

\title{Extraction of Various Five-Quark Components of the Nucleons}

\author[a]{Wen-Chen Chang} \author[a,b]{Jen-Chieh Peng}

\address[a]{Institute of Physics, Academia Sinica, Taipei 11529,
Taiwan}

\address[b]{Department of Physics, University of Illinois at
Urbana-Champaign, Urbana, Illinois 61801, USA}

\begin{abstract}
We have generalized the approach of Brodsky {\it et al.} for the
intrinsic charm quark distribution in the nucleons to the light-quark
sector involving intrinsic $\bar u, \bar d , s$ and $\bar s$ sea
quarks. We compare the calculations with the existing $\bar d - \bar
u$, $s + \bar s$, and $\bar u + \bar d - s -\bar s$ data. The good
agreement between the theory and the data allows the extraction of the
probabilities for the $|uudu\bar{u}\rangle$, $|uudd\bar{d}\rangle$,
and $|uuds\bar{s}\rangle$ five-quark Fock states in the proton. We
also calculate the $x$-dependence of the intrinsic charm after taking
into consideration the QCD evolution of the intrinsic quark
distribution.
\end{abstract}

\begin{keyword}
% keywords here, in the form: keyword \sep keyword
nucleon structure \sep five-quark states \sep intrinsic sea quark
% PACS codes here, in the form: \PACS code \sep code
\PACS 12.38.Lg \sep 14.20.Dh \sep 14.65.Bt \sep 14.65.Dw \sep 13.60.Hb
\end{keyword}

\end{frontmatter}

%\date{\today}

%\section{Introduction}

The origin of sea quarks of the nucleons remains a subject of intense
interest in hadron physics. Brodsky, Hoyer, Peterson, and Sakai
(BHPS)~\cite{brodsky80} suggested some time ago that there are two
distinct components of the nucleon sea. The first is called the
``extrinsic" sea originating from the splitting of gluons into $Q \bar
Q$ pairs. This extrinsic sea can be well described by quantum
chromodynamics (QCD). Another component of the nucleon sea is the
``intrinsic" sea which has a nonperturbative origin. In particular,
the $|u u d Q \bar Q\rangle$ five-quark Fock states can lead to the
``valence-like" intrinsic sea for the $Q$ and $\bar Q$ in the
proton. This intrinsic component is expected to carry a relatively
large momentum fraction $x$, in contrast to the extrinsic one peaking
at the small-$x$ region. Brodsky {\it et al.}~\cite{brodsky80}
proposed that the $|u u d c \bar c\rangle$ five-quark state can lead
to enhanced production of charmed hadrons at the forward rapidity
region. The CTEQ collaboration~\cite{pumplin} has examined all relevant 
hard-scattering
data and concluded that the data are consistent with a wide range of
the intrinsic charm magnitude, ranging from null to 2-3 times larger than
the estimate by the BHPS model. This suggests that more precise experimental
measurements are needed for determining the magnitude of the intrinsic
charm component.

In a recent work~\cite{chang11}, we generalized the BHPS model of the
five-quark Fock states to the light-quark sector. This work was
motivated by the expectation that the probability for the $|u u d Q
\bar Q\rangle$ Fock state is approximately proportional to $1/m_Q^2$,
where $m_Q$ is the mass of the quark $Q$~\cite{brodsky80}. Although
this $1/m_Q^2$ dependence is applicable only when the quark mass
is heavy~\cite{franz00}, 
the light five-quark states $|u u d u \bar u\rangle$, $|u u d d \bar
d\rangle$ and $|u u d s \bar s\rangle$ are likely to have
significantly larger probabilities than the $|u u d c \bar c\rangle$
state, and could be more readily observed experimentally.

By solving the Bjorken-$x$ distribution of the $\bar Q$ sea quark for
the $|u u d Q \bar Q\rangle$ five-quark state in the BHPS model
numerically, it was found~\cite{chang11} that the existing $\bar d (x)
- \bar u(x)$ and $\bar u(x) + \bar d(x) - s(x) - \bar s(x)$ data can
be well described by the calculation, provided that the QCD
evolution~\cite{dglap} of these distributions is taken into
account. Moreover, the probabilities for the $|u u d u \bar u\rangle$
and the $|u u d d \bar d\rangle$ five-quark states could also be
extracted from these data. However, the extracted values of these two
probabilities depend on the assumption adopted for the probability of
the $|u u d s \bar s\rangle$ state~\cite{chang11}.

In this paper, the previous work is extended further to determine the
probability of the $|u u d s \bar s\rangle$ five-quark state using the
recent $s(x) + \bar s(x)$ data from the HERMES
collaboration~\cite{hermes}. We found that the $s(x) + \bar s(x)$ data
in the $x>0.1$ region are quite well described by the BHPS model,
allowing the extraction of the probability of the $|u u d s \bar
s\rangle$ state. Using this probability for the $|u u d s \bar
s\rangle$ five-quark component, more precise values for the $|u u d u
\bar u\rangle$ and the $|u u d d \bar d\rangle$ states could then be
obtained from the comparison of the BHPS calculations with the $\bar d
(x) - \bar u(x)$ and $\bar u(x) + \bar d(x) - s(x) - \bar s(x)$
data. We have also examined the effect of the QCD evolution on the $x$
distribution of the intrinsic charm. In particular, we note that the
region most sensitive to intrinsic charm is shifted to lower $x$ as a
result of QCD evolution. This has implication on future searches for
intrinsic charm.

%\section{Intrinsic five-quark state of the proton}

For a $|u u d Q \bar Q\rangle$ five-quark Fock state of the proton,
the probability for quark $i$ to carry a momentum fraction $x_i$ is
given in the BHPS model~\cite{brodsky80} as
\begin{equation}
P(x_1, ...,x_5)=N_5\delta(1-\sum_{i=1}^5x_i)
[m_p^2-\sum_{i=1}^5\frac{m_i^2}{x_i}]^{-2},
\label{eq:prob5q_a}
\end{equation}
\noindent where the delta function ensures that the proton momentum is
shared among the individual constituents. $N_5$ is the normalization
factor for the five-quark Fock state, and $m_i$ is the mass of quark
$i$.  Eq.~\ref{eq:prob5q_a} was solved analytically in
Ref.~\cite{brodsky80} for the limiting case of $m_{4,5} >> m_p,
m_{1,2,3}$, where $m_p$ is the proton mass. For the more general case,
Eq.~\ref{eq:prob5q_a} can be solved numerically as discussed in
Ref.~\cite{chang11}. In particular, the $x$ distribution of $\bar Q$
in the $|u u d Q \bar Q\rangle$ state, called $P^{Q\bar Q}(x_{\bar
Q})$, can be calculated numerically.  The moment of $P^{Q\bar
Q}(x_{\bar Q})$ is defined as ${\cal P}^{Q \bar Q}_5$, namely,
\begin{equation}
{\cal P}^{Q \bar Q}_5 = \int^{1}_{0} P^{Q \bar Q}(x_{\bar Q}) dx_{\bar Q}.
\label{eq:prob5}
\end{equation}
\noindent ${\cal P}^{Q \bar Q}_5$ represents the probability of the 
$|uud Q \bar Q\rangle$ five-quark Fock state in the proton. In the limit of
$m_{4,5} >> m_p, m_{1,2,3}$, one can obtain~\cite{brodsky80} 
${\cal P}^{Q \bar Q}_5 = N_5/(3600 m_{4,5}^4)$. For the more general
case, the relation between ${\cal P}^{Q \bar Q}_5$ and $N_5$ can be
calculated numerically~\cite{chang11}.

%\section{Intrinsic $|u u d u \bar u\rangle$ and $|u u d d \bar d\rangle$}

\begin{figure}[t]
\includegraphics[width=0.5\textwidth]{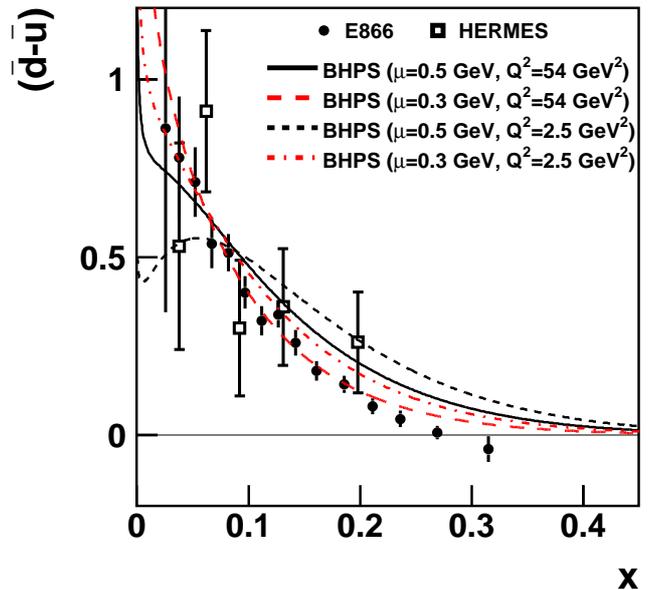}
\caption{Comparison of the $\bar d(x) - \bar u(x)$ data from
Fermilab E866 and HERMES with the
calculations based on the BHPS model. Eq.~\ref{eq:prob5q_a} and
Eq.~\ref{eq:intdbarubar1} were used to calculate the $\bar d(x) - \bar u(x)$
distribution at the initial scale. The distribution was then evolved to 
the $Q^2$ of the experiments and shown as various curves. Two different 
initial scales, $\mu = 0.5$ and $0.3$ GeV, were used for the 
E866 calculations in order to illustrate the dependence on the choice
of the initial scale.}
\label{fig_dbarubar}
\end{figure}

To compare the experimental data with the prediction based on the
intrinsic five-quark Fock state, it is necessary to separate the
contributions of the intrinsic sea quark and the extrinsic one. The
$\bar d(x) - \bar u(x)$ is an example of quantities which are free
from the contributions of the extrinsic sea quarks, since the
perturbative $g \to Q \bar Q$ processes will generate $u \bar u$ and
$d \bar d$ pairs with equal probabilities and have no contribution to
this quantity. The $\bar d(x) - \bar u(x)$ data from the Fermilab
E866 Drell-Yan experiment at the $Q^2$ scale of 54 GeV$^2$~\cite{e866}
are shown in Fig.~\ref{fig_dbarubar}. Also shown in
Fig.~\ref{fig_dbarubar} are the data obtained at a lower scale of $Q^2
= 2.5$ GeV$^2$ by the HERMES collaboration in a semi-inclusive
deep-inelastic scattering (SIDIS) experiment~\cite{hermes_sidis}.

% Figure 1

The BHPS model has a specific prediction on the shapes of the $x$
distributions for $\bar d$ and $\bar u$, since these anti-quarks
originate from the $|u u d d \bar d\rangle$ and $|u u d u \bar
u\rangle$ configurations and can be readily calculated. In the BHPS
model, the $\bar u$ and $\bar d$ are predicted to have the same
$x$-dependence if $m_u = m_d$. However, the probabilities of the $|u u
d d \bar d\rangle$ and $|u u d u \bar u\rangle$ configurations, ${\cal
P}^{d \bar d}_5$ and ${\cal P}^{u \bar u}_5$, are not known from the
BHPS model, and remain to be determined by the
experiments. Non-perturbative effects such as
Pauli-blocking~\cite{feynman} could lead to different probabilities
for the $|u u d d \bar d\rangle$ and $|u u d u \bar u\rangle$
configurations. Nevertheless the shape of the $\bar d(x) - \bar u(x)$
distribution shall be identical to those of $\bar d(x)$ and $\bar
u(x)$ in the BHPS model. Moreover, the normalization of $\bar d(x) -
\bar u(x)$ is known from the measurement of Fermilab E866 Drell-Yan
experiment~\cite{e866} as
\begin{equation}
\int^{1}_{0} (\bar d(x) - \bar u(x)) dx = 
{\cal P}^{d \bar d}_5 - {\cal P}^{u \bar u}_5 = 0.118 \pm 0.012.
\label{eq:intdbarubar1}
\end{equation}
\noindent Equation~\ref{eq:intdbarubar1} allows us to compare 
the calculations from the BHPS model with the $\bar d(x) - \bar u(x)$ data.

The $\bar d(x) - \bar u(x)$ distribution from the BHPS model is
first calculated using Eq.~\ref{eq:prob5q_a} with $m_u = m_d =
0.3$ GeV/$c^2$, and $m_p = 0.938$ GeV/$c^2$, and 
Eq.~\ref{eq:intdbarubar1} for the normalization. Since the E866 and the
HERMES data were obtained at $Q^2$ of 54 GeV$^2$ and 2.5 GeV$^2$,
respectively, it is important to evolve the $\bar d(x) - \bar u(x)$ 
distribution from the initial scale $\mu$, expected to be around the 
confinement scale, to the $Q^2$ corresponding to the data. As
$\bar d(x) - \bar u(x)$ is a flavor non-singlet parton distribution, 
its evolution from $\mu$ to $Q$ only depends on the values of 
$\bar d(x) - \bar u(x)$ at the initial scale $\mu$,
and can be readily calculated using the non-singlet evolution 
equation~\cite{dglap}. 
For the initial scale, we adopt the value of $\mu =
0.5$ GeV, which was chosen by Gl\"{u}ck, Reya, and Vogt~\cite{grv} in
the so-called ``dynamical approach" using only valence-like
distributions at the initial $\mu^2$ scale and relying on evolution to
generate the quark and gluon distributions at higher $Q^2$.

The solid and dashed curves in Fig.~\ref{fig_dbarubar} correspond to
$\bar d(x) - \bar u(x)$ calculated from the BHPS model evolved to
$Q^2=$ 54 GeV$^2$ using $\mu = 0.5$ and $0.3$ GeV, respectively. The
$x$-dependence of the E866 $\bar d(x) -\bar u(x)$ data is quite well
described by the five-quark Fock states in the BHPS model provided
that the $Q^2$-evolution is taken into consideration. It is
interesting to note that an excellent fit to the data can be obtained
if $\mu = 0.3$ GeV is chosen (dashed curve in Fig.~\ref{fig_dbarubar})
rather than the more conventional value of $\mu = 0.5$ GeV. Also shown
in Fig.~\ref{fig_dbarubar} are the calculations with the BHPS model
evolved to $Q^2 =2.5$ GeV$^2$ using $\mu = 0.5$ GeV and $\mu =0.3$
GeV. The calculations are in agreement with the HERMES data within the
experimental uncertainties.

%\section{Intrinsic $|u u d s \bar s\rangle$}

\begin{figure}[t]
\includegraphics[width=0.5\textwidth]{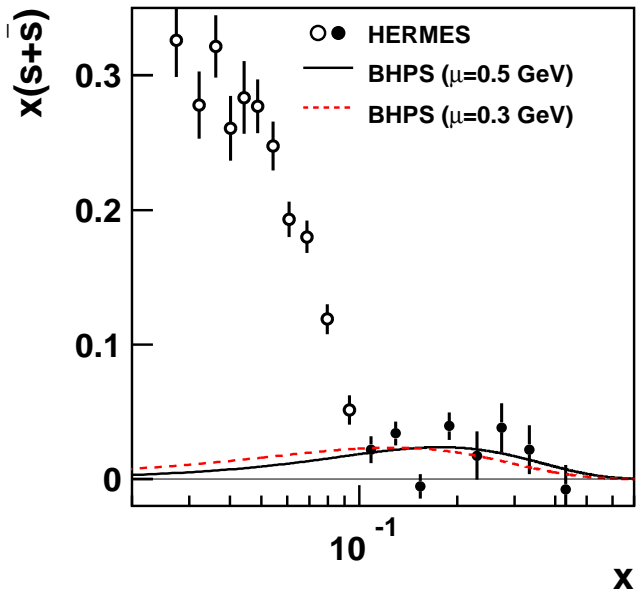}
\caption{Comparison of the HERMES $x(s(x) + \bar s(x))$ data with the
calculations based on the BHPS model. The solid and dashed curves are
obtained by evolving the BHPS result to $Q^2 = 2.5$ GeV$^2$ using $\mu
= 0.5$ GeV and $\mu = 0.3$ GeV, respectively. The normalizations of
the calculations are adjusted to fit the data at $x > 0.1$ with
statistical errors only, denoted by solid circles.}
\label{fig_ssbar}
\end{figure}

We now consider the extraction of the $|uuds \bar s\rangle$ five-quark
component from existing data. The HERMES collaboration reported the
determination of $x(s(x) + \bar s(x))$ over the range of $0.02 < x <
0.5$ at $Q^2 = 2.5$ GeV$^2$ from their measurement of charged kaon
production in SIDIS reaction~\cite{hermes}. The HERMES data, shown in
Fig.~\ref{fig_ssbar}, exhibits an intriguing feature. A rapid fall-off
of the strange sea is observed as $x$ increases up to $x \sim 0.1$,
above which the data become relatively independent of $x$. The data
suggest the presence of two different components of the strange sea,
one of which dominates at small $x$ $(x<0.1)$ and the other at larger
$x$ $(x>0.1)$. This feature is consistent with the expectation that
the strange-quark sea consists of both the intrinsic and the extrinsic
components having dominant contributions at large and small $x$
regions, respectively. In Fig. 2 we compare the data with calculations
using the BHPS model with $m_s=0.5$ GeV/c$^2$. The solid and dashed
curves are results of the BHPS model calculations evolved to $Q^2 =
2.5$ GeV$^2$ using $\mu = 0.5$ GeV and $\mu = 0.3$ GeV,
respectively. The normalizations are obtained by fitting only data
with $x>0.1$ (solid circles in Fig.~\ref{fig_ssbar}), following the
assumption that the extrinsic sea has negligible contribution relative
to the intrinsic sea in the valence region. Figure 2 shows that the
fits to the data are quite adequate, allowing the extraction of the
probability of the $|uuds \bar s\rangle$ state as
\begin{eqnarray}
{\cal P}^{s \bar s}_5 = 0.024~~(\mu = 0.5~\rm{GeV}); \nonumber \\
{\cal P}^{s \bar s}_5 = 0.029~~(\mu = 0.3~\rm{GeV}).
\label{eq:ssbar_value}
\end{eqnarray}

% Figure 2

%\section{Probabilities of intrinsic $|u u d Q \bar Q\rangle$ of light quark $Q$}

\begin{figure}[t]
\includegraphics[width=0.5\textwidth]{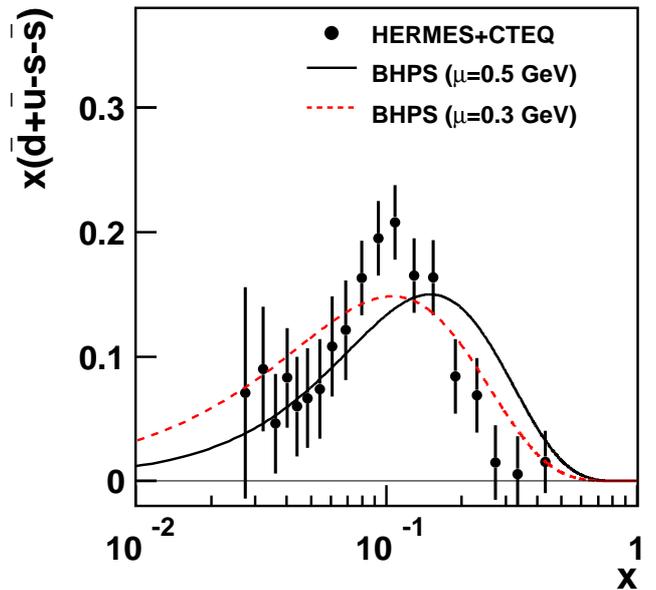}
\caption{Comparison of the $x(\bar d(x) + \bar u(x) - s(x) - \bar
s(x))$ data with the calculations based on the BHPS model. The values
of $x(s(x) + \bar s(x))$ are from the HERMES experiment~\cite{hermes},
and those of $x(\bar d(x) + \bar u(x))$ are obtained from the PDF set
CTEQ6.6~\cite{cteq}. The solid and dashed curves are
obtained by evolving the BHPS result to $Q^2 = 2.5$ GeV$^2$ using $\mu
= 0.5$ GeV and $\mu = 0.3$ GeV, respectively. The normalization of 
the calculations are adjusted to fit the data.}
\label{fig_sbar-dubar}
\end{figure}

We consider next the quantity $\bar u(x) + \bar d(x) - s(x) - \bar
s(x)$. Combining the HERMES data on $x(s(x)+ \bar s(x))$ with the
$x(\bar d(x) + \bar u(x))$ distributions determined by the CTEQ group
(CTEQ6.6)~\cite{cteq}, the quantity $x(\bar u(x) + \bar d(x) - s(x) -
\bar s(x))$ can be obtained and is shown in
Fig.~\ref{fig_sbar-dubar}. This approach for determining $x(\bar u(x)
+ \bar d(x) - s(x) - \bar s(x))$ is identical to that used by Chen,
Cao, and Signal in their recent study of strange quark sea in the
meson-cloud model~\cite{signal}.

An important property of $\bar u + \bar d - s - \bar s$ is that the
contribution from the extrinsic sea vanishes, just like the case for
$\bar d - \bar u$. Therefore, this quantity is only sensitive to the
intrinsic sea and can be compared with the calculation of the
intrinsic sea in the BHPS model. We have
\begin{eqnarray}
\bar u(x) + \bar d(x) - s(x) - \bar s(x) = \nonumber \\
P^{u \bar u}(x_{\bar u}) + 
P^{d \bar d}(x_{\bar d}) - 2 P^{s \bar s}(x_{\bar s}).
\label{eq:udssbar_p5}
\end{eqnarray}
\noindent We can now compare the $x(\bar u(x) + \bar d(x) - s(x) -
\bar s(x))$ data with the calculation using the BHPS model. Since
$\bar u + \bar d - s - \bar s$ is a flavor non-singlet quantity, we
can readily evolve the BHPS prediction to $Q^2 =2.5$ GeV$^2$ using
$\mu = 0.5$ GeV and the result is shown as the solid curve in
Fig.~\ref{fig_sbar-dubar}. It is interesting to note that a better fit
to the data can again be obtained with $\mu = 0.3$ GeV, shown as the
dashed curve in Fig.~\ref{fig_sbar-dubar}.

% Figure 3

From the comparison between the data and the BHPS calculations shown
in Figs.~\ref{fig_dbarubar}-\ref{fig_sbar-dubar}, we can determine
the probabilities for the $|u u d u \bar u\rangle$, $|u u d d \bar
d\rangle$, and $|u u d s \bar s\rangle$ configurations as follows:
\begin{eqnarray}
{\cal P}^{u \bar u}_5 = 0.122;~~{\cal P}^{d \bar d}_5 = 0.240;~~{\cal P}^{s \bar s}_5 = 0.024~~ \nonumber \\
(\mu = 0.5~\rm{GeV})
\label{eq:uds_value_a}
\end{eqnarray}
\noindent or
\begin{eqnarray}
{\cal P}^{u \bar u}_5 = 0.162;~~{\cal P}^{d \bar d}_5 = 0.280;~~{\cal P}^{s \bar s}_5 = 0.029~~ \nonumber \\
(\mu = 0.3~\rm{GeV})
\label{eq:uds_value_b}
\end{eqnarray}
\noindent depending on the value of the initial scale $\mu$. It is
remarkable that the $\bar d(x) - \bar u(x)$, the $s(x) + \bar s(x)$,
and the $\bar d(x) + \bar u(x) - s(x) - \bar s(x)$ data not only allow
us to check the predicted $x$-dependence of the five-quark Fock
states, but also provide a determination of the probabilities for
these states.

%\section{Discussion and Conclusion}

Equations 6 shows that the combined probability for proton to be in
the $|uud Q \bar Q\rangle$ states is around 40\%. It is worth noting that
an earlier analysis of the $\bar d - \bar u$ data in the meson cloud model
concluded that proton has $\sim$60\% probability to be in the three-quark
bare-nucleon state~\cite{szczurek}, in qualitative agreement with the
finding of this study.  A significant feature of the present work is 
the extraction of the $|uud s \bar s\rangle$ component, which would be
related to the kaon-hyperon states in the meson cloud model. It is also
worth mentioning that in the BHPS model the $|uud Q \bar Q\rangle$ states
have the same contribution to the proton's magnetic moment as the 
$|uud\rangle$ three-quark state, since $Q$ and $\bar Q$ in the 
$|uud Q \bar Q\rangle$ states have no net magnetic moment. Therefore, 
the good description of the nucleon's magnetic moment by the constituent
quark model is preserved even with the inclusion of a sizable five-quark
components in the BHPS model.

We note that the probability for the $|uud s \bar s\rangle$ state is
smaller than those of the $|uud u \bar u\rangle$ and the $|uud d \bar
d\rangle$ states. This is consistent with the expectation that the
probability for the $|uud Q \bar Q\rangle$ five-quark state is roughly
proportional to $1/m^2_Q$~\cite{brodsky80,franz00}. One can then 
estimate that the probability
for the intrinsic charm from the $|uud c \bar c\rangle$ Fock state,
${\cal P}^{c \bar c}_5$ to be roughly 0.01. This is also consistent
with an estimate based on the bag model~\cite{donoghue77}, as well as
with an analysis of the EMC charm-production data~\cite{harris96}. 
Figure 4 shows
the $x$ distribution of intrinsic $\bar c$ calculated with the BHPS
model using 1.5 GeV/c$^2$ for the mass of the charm quark. Also shown
in Fig. 4 is the calculation which evolve the BHPS calculation from
the initial scale, $\mu=0.5$ GeV, to $Q^2=75$ GeV$^2$, the largest $Q^2$ scale reached by EMC~\cite{emc}. It is interesting to
note that the intrinsic charm contents at the large $x$ ($x>0.3$)
region are drastically reduced when $Q^2$ evolution is taken into
account. Figure 4 suggests that the most promising region to search
for evidence of intrinsic charm could be at the somewhat lower $x$
region $(0.1 < x < 0.4)$, rather than the largest $x$ region explored
by previous experiments. It is worth noting that we adopt the simple
assumption that the initial scale is the same for all five-quark states.
It is conceivable that the initial scale for intrinsic charm is
significantly higher due to the larger mass of the charmed quark.
The dashed curve shows the $x$ distribution of intrinsic $\bar c$
at $Q^2 = 75$ GeV$^2$ when the initial scale is set at $\mu = 3$ GeV,
corresponding to the threshold of producing a pair of charmed quarks.
As expected, the shape of the intrinsic $\bar c$ $x$ distribution
becomes similar to that of the BHPS model.

% Figure 4
\begin{figure}[t]
\includegraphics[width=0.5\textwidth]{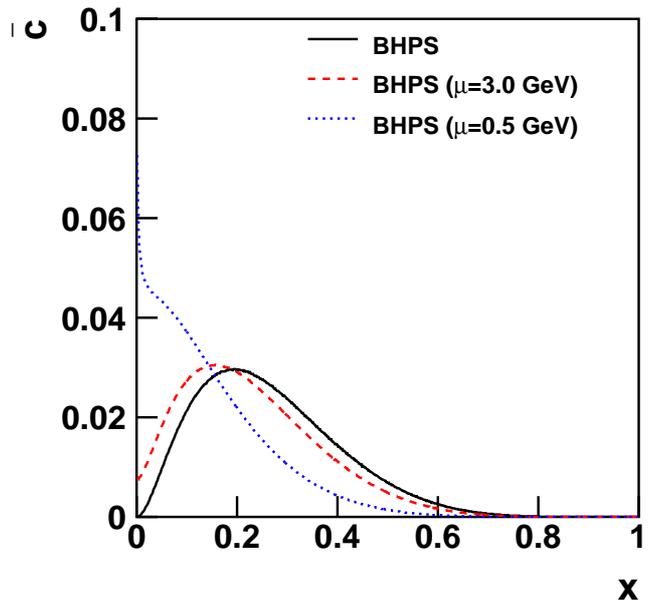}
\caption{Calculations of the $\bar c(x)$ distributions based on 
the BHPS model. The solid curve corresponds to
the calculation using Eq.~\ref{eq:prob5q_a} and the dashed and dotted curves are obtained by evolving the BHPS result to 
$Q^2 = 75$ GeV$^2$ using
$\mu = 3.0$ GeV, and $\mu = 0.5$ GeV, respectively. 
The normalization is
set at ${\cal P}^{c \bar c}_5 = 0.01$.}
\label{fig_cbarr}
\end{figure}

In conclusion, we have generalized the existing BHPS model to the
light-quark sector and compared the calculation with the $\bar d -
\bar u$, $s + \bar s$, and $\bar u + \bar d - s - \bar s$ data. The
qualitative agreement between the data and the calculations provides
strong support for the existence of the intrinsic $u$, $d$ and $s$
quark sea and the adequacy of the BHPS model. This analysis also led
to the determination of the probabilities for the five-quark Fock
states for the proton involving light quarks only. This result could
guide future experimental searches for the intrinsic $c$ quark sea or
even the intrinsic $b$ quark sea~\cite{ma}, which could be relevant for
the production of Higgs boson at LHC energies~\cite{brodsky06}.

\section*{Acknowledgments}

We acknowledge helpful discussion with Stan Brodsky, Gerry Garvey, Don
Geesaman, Bo-Qiang Ma, Tony Thomas, and Fan Wang. This work was
supported in part by the National Science Council of the Republic of
China and the U.S. National Science Foundation. One of the authors
(J.P.) thanks the members of the Institute of Physics, Academia Sinica
for their hospitality.

\end{document}